\documentclass[11pt,reqno]{article}
\usepackage{amsmath}
\usepackage{amsfonts}
\usepackage{amssymb}
\usepackage{hyperref}
\usepackage{latexsym}
\usepackage[dvips]{graphicx}
\usepackage{epsf}
\usepackage{color}

\allowdisplaybreaks \numberwithin{equation}{section}

\newcommand{\be}{\begin{equation}}
\newcommand{\ee}{\end{equation}}
\newcommand{\bea}{\begin{eqnarray}}
\newcommand{\eea}{\end{eqnarray}}

\newcommand{\f}{\frac}
\newcommand{\p}{\partial}
\newcommand{\na}{\nabla}
\newcommand{\Tr}{{\rm Tr}}

\let\a=\alpha \let\b=\beta  \let\g=\gamma  \let\d=\delta
\let\z=\zeta       \let\k=\kappa \let\l=\lambda
\let\m=\mu    \let\n=\nu    \let\x=\xi

\let\G=\Gamma    \let\L=\Lambda 
    \let\Si=\Sigma     
  \let\eps=\epsilon

\newcommand{\gt}{\tilde{g}}

\newcommand{\cD}{\mathcal{D}}

\newcommand{\cG}{\mathcal{G}}

\newcommand{\pr}[1]{ {}^{(#1)}\hspace{-0.05cm} }

\DeclareFontFamily{OT1}{pzc}{}     
\DeclareFontShape{OT1}{pzc}{m}{it}{<-> s * [1.25] pzcmi7t}{}
\DeclareMathAlphabet{\mathpzc}{OT1}{pzc}{m}{it}

\begin{document}

\thispagestyle{empty}
\begin{flushright} \small
AEI-2013-264
\end{flushright}
\bigskip

\begin{center}
 {\LARGE\bfseries  One-loop renormalization in\\ [1.5ex] 
a toy model of Ho\v{r}ava-Lifshitz gravity  
}
\\[10mm]
{\large Dario Benedetti${}^a$ and Filippo Guarnieri${}^{a,b}$}
\\[3mm]
{\small\slshape
${}^a$ Max Planck Institute for Gravitational Physics (Albert Einstein Institute), \\
Am M\"{u}hlenberg 1, D-14476 Golm, Germany \\ 
${}^b$ Dipartimento di Fisica, Universit$\grave{a}$ degli Studi di Roma Tre and\\ 
INFN sezione di Roma Tre, Via della Vasca Navale 84, I-00146 Rome, Italy\\
\vspace{.3cm}
 {\upshape\ttfamily dario.benedetti@aei.mpg.de, filippo.guarnieri@aei.mpg.de} }
\end{center}
\vspace{5mm}

\hrule\bigskip

\centerline{\bfseries Abstract} \medskip
\noindent

We present a one loop calculation in the context of Ho\v rava-Lifshitz gravity. Due to the complexity of the calculation in the full theory we focus here on the study of a toy model, namely the conformal reduction of the $z=2$ projectable theory in $2+1$ dimensions. For this value of the dimension there are no gravitons, hence the conformal mode is the only physical degree of freedom, and thus we expect our toy model to lead to qualitatively correct answers regarding the perturbative renormalization of the full theory. 
We find that Newton's constant (dimensionless in Ho\v rava-Lifshitz gravity) is asymptotically free. However, the DeWitt supermetric approaches its Weyl invariant form with the same speed and the effective interaction coupling remains constant along the flow.
In other words, the would-be asymptotic freedom associated to the running Newton's constant is exactly balanced by the strong coupling of the scalar mode as the Weyl invariant limit is approached.
We conclude that in such model the UV limit is singular at one loop order, and we argue that a similar phenomenon can be expected in the full theory, even in higher dimensions.

\bigskip
\hrule\bigskip
\newpage
\tableofcontents

\section{Introduction}

The standard quantum field theory approach to a perturbative quantization of gravity is notoriously hindered by the clash between  renormalizability and unitarity.
It was suggested by Ho\v{r}ava \cite{Horava:2009uw} that the two could be reconciled if we are ready to give up another pillar of standard quantum field theory, Lorentz invariance.
By introducing a preferred spacetime slicing, and constructing an action with sufficiently higher-order spatial derivatives, but with at most two time derivatives, we can obtain a power-counting renormalizable theory of gravity. Such models are now known as Ho\v{r}ava-Lifshitz gravity, or HL gravity for brevity, and they have been the subject of much study. Despite the obvious drawback of lost Lorentz invariance, which in particular forces such models to face big observational challenges and fine tuning problems \cite{Iengo:2009ix},\footnote{A phenomenologically viable scenario that could avoid such fine-tuning problems has been proposed in \cite{Pospelov:2010mp}.} the appealing feature of a renormalizable model of gravity in the usual sense has made HL gravity an intensely studied topic.\footnote{Other motivations are found for example in cosmology \cite{Mukohyama:2010xz}, in the relation to causal dynamical triangulations \cite{Horava:2009if,Ambjorn:2010hu,Anderson:2011bj}, and in the possibility of using HL gravity as a holographic dual to non-relativistic theories \cite{Griffin:2012qx,Janiszewski:2012nb}.}
Oddly, the renormalization properties of HL gravity, arguably their main motivation, are to date their least explored feature.\footnote{With of course few important exceptions \cite{Orlando:2009en,Giribet:2010th,Nesterov:2010yi,Nakayama:2012sn,Rechenberger:2012dt,Contillo:2013fua}.}
Almost nothing is known about loop corrections to the HL action, and a full proof of renormalizability is still missing. 
In particular, we do not know yet whether the theory is asymptotically free or if it suffers from triviality. Neither do we know whether the theory flows towards general relativity in the infrared.

The reasons for the scarcity of results on the renormalization of HL gravity are easily identifiable in the complexity of the required calculations, due to the lack of covariance (or equivalently the need to introduce a unit timelike vector \cite{Jacobson:2010mx}), as well as to the large number of terms present in the action of the most general model, i.e. the non-projectable model without detailed balance \cite{Blas:2009qj}.
A very common strategy in trying to make progress in similar situations is to identify some essential features of the model we aim at, and study a simplified version of it in which such essential features are maintained while most of the complications are set aside.

One first simplification which we will adopt here, is to reduce the number of spacetime dimensions. In classical general relativity, four is the smallest number of dimensions in which the theory has propagating degrees of freedom, but three dimensional quantum gravity has nevertheless been a very active field of research, due to the fact that it shares many problematics with its higher-dimensional version \cite{Carlip:2004ba}.
In the case of HL gravity, the three dimensional case might be even more interesting, because while gravitons are still absent, the new scalar degree of freedom associated to the breaking of full diffeomorphism invariance is still present, thus allowing us to concentrate on it without the distraction from the gravitons. 
In fact, lower dimensional models of HL gravity have already received some attention \cite{Horava:2008ih,Benedetti:2009ge,Sotiriou:2011dr,Ambjorn:2013joa}.
However, it turns out that in order to study the running of all the couplings at one loop order, even in three dimensions, and for the simple $z=2$ projectable model, some technical annoyances persist.
In order to simplify matters as much as possible, and to get a glimpse over the questions we raised above about renormalization, we will adopt one second main simplification, i.e. after having gauge-fixed lapse and shift, we will quantize only the conformal mode of the spatial metric.
A similar {\it conformal reduction} has also been widely adopted as a toy model in other contexts. One example, close to our setting, is the use of conformally reduced gravity models in studying the asymptotic safety scenario \cite{Reuter:2008wj,Machado:2009ph,Daum:2009qe,Bonanno:2012dg, Bonanno:2013dja}.
It is actually somewhat surprising that anything can be learned from such a reduction in the case of standard isotropic gravity, as in general relativity the scalar mode is not a propagating degree of freedom. Quite on the contrary, in the case of three-dimensional HL gravity, the scalar mode is the only physical degree of freedom, as gravitons are absent and the longitudinal modes are killed by the constraints, and therefore we might expect the conformally reduced model to be much closer to the full theory.

We will derive the form of the divergences arising in the effective action of our toy model at one loop, and translate them into beta functions for the renormalization group running of the dimensionless couplings. We will see that while the running of Newton's constant might suggest a realization of asymptotic freedom, the situation is complicated by the running of the DeWitt supermetric, leading to an effective coupling which remains finite at all scales.

We will begin in Sec.~\ref{Sec:action} by presenting the model, while in Sec.~\ref{Sec:bfm} we will introduce the background field splitting and illustrate the peculiarities of the field content in three dimensions. In Sec.~\ref{Sec:symm} we will discuss the symmetries of the model, and we will introduce gauge-fixing and ghosts for the quantization procedure.
Later, in Sec.~\ref{Sec:oneloop} we will explain the general one-loop algorithm and introduce the effective coupling.
Finally, in Sec.~\ref{Sec:beta} our main calculation and results will be detailed, followed by a discussion of the results in Sec.~\ref{Sec:concl}.

\section{The action}
\label{Sec:action}

We assume a spacetime topology $\mathbb{R}\times\Si$, with $\Si$ a closed two-dimensional manifold,
and we choose Euclidean signature for the spacetime metric, which we will decompose according to the standard ADM splitting, keeping the spacetime nomenclature despite the Euclidean signature.

Following \cite{Horava:2009uw}, a HL gravity theory is constructed by giving mass dimension $-z$ to the time coordinate, $[t]=-z$, and standard dimension to the spatial coordinates, $[x^i]=-1$, and by building an action invariant under foliation-preserving diffeomorphisms. Power-counting renormalizability in $d+1$ dimensions is obtained by choosing $z=d$, and by including in the bare Lagrangian all the possible local operators compatible with the symmetries and with mass-dimension up to $2z$. The latter condition, together with the dimensions assigned to time, automatically implies that no more than two time derivatives appear in the action, thus preserving unitarity, at least in the naive sense. At the same time the inverse propagator now contains up to $2z$ powers of spatial derivatives, thus improving convergence of the loop integrals.
One obtains a super-renormalizable theory for $z>d$, and a non-renormalizable one for $z<d$. We are interested in the just renormalizable case in $2+1$ dimensions, hence we will consider the theory with $z=2$. Such model was first considered in \cite{Horava:2008ih}, but with detailed balance condition for the potential, which for $d=2$ leads to no potential at all. Here we will study the case without detailed balance, which was also considered in \cite{Benedetti:2009ge,Sotiriou:2011dr}.

There are two main versions of HL gravity, respectively known as projectable and non-projectable version.
The projectable version is characterized by a spatially constant lapse function, $N=N(t)$, and its most generic $z=2$ action reads
\be\label{action}
S =  \frac{2}{\k^2} \, \int d t\, d^2 x N \sqrt{g} \, \left\{  \l\, K^2 -K_{ij}K^{ij} -2\, \L + c\, R +\g\, R^2  \right\}\, ,
\ee
where $g$ is the determinant of the spatial metric, $R$ its Ricci scalar, $N$ the lapse function, $K_{ij}$ the extrinsic curvature of the leaves of the foliation, and $K$ its trace. The coupling $\k^2$ is proportional to Newton's constant, and $\L$ is the cosmological constant, while $\l$ and $\g$ characterize the deviations from full diffeomorphism invariance ($\l=1$ and $\g=0$ corresponding to general relativity in 2+1 dimensions\footnote{Note that we have chosen the sign of the kinetic term in such a way that the quadratic action for the conformal mode has the correct sign for $\l=1$, unlike in general relativity. This makes sense in $2+1$ dimensions because there are no gravitons.}).
In particular, $\l$ defines a one-parameter family of deformed DeWitt supermetrics
\be
\cG^{ijkl} = \f12 \left(g^{ik}g^{jl}+g^{il}g^{jk} \right) - \l\, g^{ij}g^{kl} \, ,
\ee
with $\l=1$ being the standard case, and $\l=\f12$ being the Weyl invariant one \cite{Horava:2008ih,Griffin:2011xs}.

In the non-projectable version, the restriction on the lapse is lifted, and the action can contain many more terms \cite{Sotiriou:2011dr}:
\be
\begin{split}
S =  \int d t\, d^2 x N \sqrt{g} & \left\{ \frac{2}{\k^2}\left(   \l\, K^2 -K_{ij}K^{ij} -2\, \L + c\, R    +\g\, R^2 \right) + c_1\, D^2\, R \right. \\
& \left. + c_2\,  a_i\, a^i + c_3\, (a_i\, a^i)^2 + c_4\, R\, a_i\, a^i + c_5\,  a_i\,  a^i\, D^j\,  a_j \right. \\
& \left. + c_6\, (D^j\, a_j)^2 + c_7\, (D_i\, a_j)(D^i\, a^j)\right\}\, .
\end{split}
\ee
Here  $a_i=D_i \ln N$ is the acceleration vector and $D_i$ the spatial covariant derivative.

The non-projectable version is clearly more demanding at a technical level, in particular from a renormalization group point of view, as even in the simplified setting of $2+1$ dimensions we have twelve couplings to take care of.
For such reason, we will in the following restrict ourselves to the projectable theory \eqref{action},
in which case, as a consequence of the Gauss-Bonnet theorem, we have also the simplification
\be \label{euler}
\int dt\, d^2x N\sqrt{g} R  =  \int dt\, N \int d^2x \sqrt{g} \pr{2}R = 4\, \pi\, \chi\, \int dt\, N \, ,
\ee
with $\chi$ the Euler characteristic of the spatial manifold $\Si$.

\section{Background field method and metric decomposition}
\label{Sec:bfm}
For our one-loop calculation we will make use of the background field method, which entails the linear splitting
\be \label{fluctuations}
g_{ij} \to g_{ij} + \epsilon\, h_{ij}\, ; \quad N \to N + \epsilon\, n\, ; \quad N_i \to N_i + \epsilon\, n_i\, ,
\ee
where $\{h_{ij},n,n_i \}$ are the quantum fluctuations, $\{g_{ij},N,N_i \}$ the background fields and $\epsilon$ is a perturbative parameter which we will set at a later stage. The background fields are in principle generic and off-shell, however, for practical purposes it suffices to choose a background that will allow us to discern the invariants of interest. In our case, it will be enough to consider a generic $g_{ij}$ and to restrict $N=1$ and $N_i=0$.

Concerning the fluctuation fields, it is convenient to use the trace-traceless decomposition
\be \label{T-dec}
h_{ij} = \hat{h}_{ij}+ \frac{1}{2} g_{ij} h\, ,
\ee
with $g^{ij}\hat{h}_{ij}=0$.
In general dimension, the traceless metric fluctuation $\hat{h}_{ij}$ can be further decomposed in transverse and longitudinal components, but in two dimensions it is well known that transverse traceless tensors form a finite dimensional vector space. In particular, on a closed manifold of genus $\mathfrak{g}$ there are  precisely $(6\mathfrak{g}-6)$ independent transverse traceless tensors for $\mathfrak{g}>1$, just two for $\mathfrak{g}=1$, and no such tensors for $\mathfrak{g}=0$.
In other words, we just recalled the well-known fact that any metric on a 2-dimensional manifold is conformal to a diffeomorphism-equivalent class of constant curvature metrics:
\be \label{conf-g}
g_{ij} = e^{2\phi} \gt_{ij} \, .
\ee
Here $\gt_{ij}$ is a metric of constant curvature, and the ensemble of such metrics modulo diffeomorphism is known as the moduli space of the manifold, which has the same dimension as the vector space discussed above, which actually is the cotangent space at $\gt_{ij}$ to the moduli space.
Hence, once we fix the topology, the metric $\gt_{ij}$ carries only gauge degrees of freedom plus a finite number of global degrees of freedom. 
We will forget about the latter in what follows, a safe way to do that being of course to choose spherical topology for the spatial slices.

The two decompositions (\ref{fluctuations}-\ref{T-dec}) and \eqref{conf-g} obviously coincide at the linear level, upon the identification $\phi=h/4$, while at higher orders they lead to inessential differences in the off-shell effective action.
The approximation we will employ in the following consists in discarding all the quantum fluctuations associated to the metric $\gt_{ij}$, which then will be treated as a background quantity, or equivalently, in discarding the traceless fluctuations $\hat{h}_{ij}$.

\section{Symmetries and gauge fixing}
\label{Sec:symm}
The action (\ref{action}) is invariant under foliation-preserving diffeomorphisms, i.e. it is invariant under the coordinate reparametrization
\bea \label{diffeos}
& x^i & \to \;\; x^i + \z^i(\vec{x},t) \, ,\\
& t  &\to \;\; t + \z(t)\, .
\eea
At leading order, the transformations of the fluctuation fields are (dots stand for time derivatives)
\bea
& h_{ij} & \to \;\; h_{ij} + D_i\, \z_j + D_j\, \z_i + \z\, \dot g_{ij} \, ,\\
& n_i  & \to \;\; n_i + g_{ij}\, \dot\z^j +\z^j\, D_j\, N_i + N_j\, D_i\, \z^j + \dot \z\,  N_i + \z\, \dot N_i \, ,\\
& n & \to \;\; n + \dot\z\, N + \z\, \dot N + \z^j D_j N \, ,
\eea
and on a background such that $N=1$ and $N_i=0$, they simplify to
\bea
& h_{ij} & \to \;\; h_{ij} + D_i\, \z_j + D_j\, \z_i + \z\, \dot g_{ij} \, , \\
& n_i  & \to \;\; n_i + g_{ij}\, \dot\z^j \, , \\
& n & \to \;\; n + \dot\z \, . 
\eea

We can use a time-dependent diffeomorphism to gauge-fix $n=n_i=0$.
There is in this case a residual symmetry, corresponding to time-independent spatial diffeomorpishms $\z^i=\z^i(\vec{x})$,
which could be fixed by a de Donder-type gauge fixing on a single slice.
A standard canonical analysis \cite{Horava:2008ih} shows that the constraints of the theory preserve such gauge fixing under time evolution, thus killing the longitudinal components of the metric fluctuations, and leaving us with only the scalar mode.
However, in a correct one-loop path integral quantization, the longitudinal modes should be integrated over without restrictions (at most just imposing the single-slice gauge-fixing as in \cite{Das:2004qk}). 
Our conformal reduction will consist in not performing such functional integration, thus freezing the longitudinal modes as if they had been eliminated by the constraints.

In order to implement the gauge condition we add the gauge-fixing action
\be
S_{gf} = \f{1}{2\, \a^2} \int dt\, N  \int d^2 x\sqrt{g}\, n^2+\f{1}{2\, \b^2} \int dt\,  \int d^2 x\sqrt{g} \, n_i \, n^i\, ,
\ee
and take the limit $\a\to 0$ and $\b\to 0$, which leads to a complete decoupling of $n$ and $n_i$.

Since the fluctuations of lapse and shift transform linearly in the time derivative, the Fadeev-Popov operator reads $\mathcal{M}=\p_t$.
In order to avoid problems inherent to the non positivity of such an operator we employ for the ghost sector the square root of the determinant of the squared Fadeev-Popov operator, namely $\sqrt{\det(-\mathcal{M}^2)}$, which also leads to better properties under the RG flow \cite{Benedetti:2011ct}.
The corresponding ghost action is then
\be
S_{gh} = \int dt\, N \int d^2 x\sqrt{g}\, \Big\{ \bar c\, \p_t^2\, c + \bar c_i\, \p_t^2\, c^i  + b\, \p_t^2\, b + b_i\, \p_t^2\, b^i  \Big\}  \, ,
\ee
being $c_i$ and $c$ Grassmannian complex fields, and $b_i$ and $b$ real bosonic fields. 
The limit $\a\to 0$ and $\b\to 0$ can be performed at the level of the second variation of the action, after the rescaling $n\to\a\, n$ and $n_i\to \b\, n_i$. It is clear that in such limit the fields $n$, and $n_i$ will only survive in the gauge-fixing term, and we can set them to zero when writing the variation of $S$.
The gauge-fixing action is clearly non-dynamical and its integration in the path integral will only give an ultralocal contribution to the action (proportional to $\d^{(3)}(0)$) which we do not keep track of. Concerning the ghosts, they will produce a determinant of $-\p_t^2$ to some power, which can only contribute to the renormalization of the cosmological constant term.

\section{One-loop setup}
\label{Sec:oneloop}
We want to evaluate the one-loop beta functions of the dimensionless coupling $\k$, $\lambda$ and $\gamma$, in order to study their renormalization group flow, and determine whether the theory is asymptotically free or not.
The one-loop effective action can be written as\footnote{Occasionally we display Planck's constant $\hslash$ as a loop expansion parameter.}
\be\label{oneloop}
\G = S_{tot} + \hslash\, S^{1-loop} + \mathcal{O}(\hslash^2), \hspace{1cm} S^{1-loop} = \f{1}{2} {\rm STr} \ln (S_{tot}^{(2)}) \, ,
\ee
where
\be
S_{tot} = S+ S_{gf}+S_{gh}\, ,
\ee
$S^{(2)}$ indicates the second functional derivative respects to the fields and STr is a supertrace (it includes a factor two for complex fields and a factor minus for Grassmann fields).

As usual, $S^{1-loop}$ will contain some UV divergences, which, being the theory renormalizable, we will be able to absorb in a renormalization of the bare couplings. The dependence of the renormalized couplings upon the renormalization scale will determine the beta functions.


The first step of the one-loop calculation is the evaluation of the second functional derivative of the action. To that end, we use the splitting \eqref{fluctuations}, under which the action decomposes as
\be
S[g_{ij} + \epsilon\,\, h_{ij}] = S[g_{ij}] + \epsilon\, \d S[g_{ij}; h_{ij}] + \epsilon^2\, \d^2 S[g_{ij}; h_{ij}] + \mathcal{O}(\epsilon^3) \, .
\ee
$S^{(2)}[g_{ij}] = \d^{(2)} S/\d h_{kl}\d h_{mn} {}_{|h=0}$ can easily be read off from  $\d^2 S[g_{ij}; h_{ij}]$ by stripping off the fluctuation fields.
As we already discussed, we will use the decomposition \eqref{T-dec} and discard the traceless contributions $\hat h_{ij}$, thus having simply  $h_{ij}=\f12 g_{ij} h$.
Expanding up to the second order in the fluctuations, we first note that in $d=2$ the variation of the metric determinant
\be
\sqrt{g} \to \sqrt{g}\, \left(1 + \epsilon\, \f12\, h + \mathcal{O}(\epsilon^3) \right) \, ,
\ee
has no part which is quadratic in the trace mode, and thus the bare cosmological constant will not enter in the one-loop correction of the action.
And due to \eqref{euler}, also the coupling $c$ in \eqref{action} will not appear in $S^{1-loop}$.

Finally, as we are not interested here in discussing the renormalization of the cosmological constant, and as the gauge-fixing and ghost term can only contribute to that, we will forget both about the lapse and shift fluctuations as well as about the ghosts.\footnote{Note that this is not an approximation: we have discussed the gauge-fixing and ghosts in Sec.~\ref{Sec:symm} precisely in order to show that they cannot contribute to the renormalization of the dimensionless couplings.}
We are thus left with a second variation depending only on the trace mode, namely
\be \label{tot-var}
\d^2 S[g_{ij}; h_{ij}] =  \frac{1}{2 \k^2} \int d t\,  d^2 x\,  \sqrt{g}\, \left\{  (\l-\f12)\, (\p_t h)^2 +\g \,  h (D^4 + 2\, R\, D^2 + R^2) h  \right\} \, .
\ee
When perturbatively quantizing general relativity, the perturbative expansion parameter $\eps$ is chosen to be equal to $\k$, so that the kinetic term for the graviton be canonically normalized.
In the present case we see that such choice is not enough, as the operator in \eqref{tot-var} depends on the two couplings $\l$ and $\g$, and there is no choice by which we could remove both of them. We should notice however that from a canonical point of view what should be normalized to one half is really the coefficient of $(\p_t h)^2$, all the rest being part of the potential. 
Restricting our analysis to the case $\l > \f12$ (for $\l<\f12$ the operator has the wrong sign, we should start again from \eqref{action} and flip the signs of the extrinsic curvature terms), we thus conclude that the effective perturbative coupling is
\be\label{eps}
\epsilon = \frac{\k}{(\l - \f12)^{1/2}} \, .
\ee
Absorbing $\eps$ into the second variation, and integrating by parts, equation \eqref{tot-var} can now be rewritten as
\be\label{tot-var-conf2}
\d^2 S = \frac{1}{2} \int d t\,  d^2 x  \sqrt{g} \, h\, \mathcal{D} \, h   \, ,
\ee
being 
\be\label{opdif}
\mathcal{D} =   - \frac{1}{\sqrt{g}}\, \p_t \sqrt{g}\, \p_t  +\frac{\gamma}{\l - \f12} \, (D^2 + R)^2 \, .
\ee
%

\section{Divergences and beta functions}
\label{Sec:beta}

The supertrace in \eqref{oneloop} reduces in our case to a single trace over the conformal mode of the spatial metric, which we will evaluate by means of a heat kernel expansion. First, we regulate the trace of the logarithm by rewriting it as\footnote{A more rigorous procedure for regularizing the functional trace would consist in using a zeta function regularization \cite{Hawking:1976ja}, however, as the final result is the same, we stick here to this more simplistic regularization scheme.}
\be \label{schwinger}
S^{1-loop} = \frac{1}{2}\,\text{Tr}\ln(\mathcal{D}) = - \frac{1}{2}\, \int_{\f{1}{\L^4}}^{+\infty} \frac{ds}{s} \, \text{Tr}\,e^{- s \mathcal{D}} \, ,
\ee
being $\mathcal{D} \equiv \frac{\delta^{2}S}{\delta h\delta h}$ the operator \eqref{opdif}, $s$ a proper time variable, and $\L$ a UV cutoff of mass dimension one (note that $[s]=-4$ due to the unusual mass-dimension of the time coordinate), not to be confused with the cosmological constant, which from now on will not appear anymore.
If the operator $\mathcal{D}$ has zero or negative modes, then expression \eqref{schwinger} will need also an IR cutoff on the upper extreme of integration.

The integrand $e^{- s\,\mathcal{D}}$ can be considered as the diagonal part of an operator
\be
\mathcal{H}(x,x',s; \mathcal{D}) = <x| \,e^{- s \mathcal{D}}\, |x'> \, ,
\ee
which satisfies the heat equation
\be
(\p_s +  \mathcal{D}) \mathcal{H} = 0 \, ,
\ee
with boundary condition
\be
\lim_{s\to0^+}\, \mathcal{H}(x,x',s;  \mathcal{D}) = \frac{1}{\sqrt{g}}\delta^{2}(x - x')\, .
\ee
A well known feature of the heat kernel is that it admits in the limit $s\to0^+$ an expansion series in powers of $s$, which in the present case reads
\be \label{HK-exp}
\mathcal{H}(x,x,s; \mathcal{D}) = \sum_{n=0}^{\infty} \, s^{\frac{n}{2}-1}\, a_n(x;\mathcal{D}) \, ,
\ee
the $a_n$ coefficients being scalars built out of geometric tensors and their derivatives.
Plugging \eqref{HK-exp} into \eqref{schwinger}, and exchanging sum and integral, we immediately find that for $n>2$ we can safely take the $\L\to\infty$ limit, and that all the UV divergences are contained in the first three terms of the expansion. By simple dimensional analysis we expect logarithmic divergences proportional to $a_2$, and we expect the latter to be a linear combination of the squares of the intrinsic and extrinsic curvatures of the spatial slices.

\subsection{Heat kernel expansion}

As a result of the heat kernel expansion, we write
\be
\label{trace-expansion}
\begin{split}
\f12 \, \Tr \ln(\mathcal{D}) =& - \f12 \,\int_{\frac{1}{\Lambda^4}}^{\frac{1}{\m^4}}\, \frac{ds}{s}\, \Tr \,e^{-s \, \mathcal{D}} = \\
& - \f12\, \int_{\frac{1}{\Lambda^4}}^{\frac{1}{\m^4}}\, \frac{ds}{s^2}\, \int dt\, d^2x \sqrt{\hat g}\, \left \{a_0 + s^\f12 \,a_1  + s \, a_2 +  \mathcal{O}(s^\f32)\right\} \, ,
\end{split}
\ee
where we have introduced also an IR cutoff $\m$ on the proper time integral, which in the Wilsonian picture plays the role of a renormalization scale.

Whereas in the isotropic case the $a_n$ coefficients of the corresponding heat kernel expansion have been worked out by many different means and for many different operators, very little is available about the anisotropic case. Luckily, for the case at hand we can take advantage of the computations done in  \cite{Baggio:2011ha}.
In fact, we can recognize that the action \eqref{tot-var-conf2} is almost the same as the one considered in that work, the only differences (beside our background choice $N=1$ which is unimportant) being the replacement $D^2\to D^2+R$ and the presence of the coupling $\g/(\l-\f12)$, 
both of which are easily taken care of.

Concerning the presence of the coupling, we can simply notice that it can be dealt with by introducing the auxiliary spatial metric
\be\label{aux}
\hat g_{ij} = \left( \frac{\l - \f12}{\g}\right)^{\f12}\, g_{ij} \, ,
\ee
so that \eqref{tot-var-conf2} now reads
\be
\d^2 S =  \frac{1}{2}\,\left( \frac{\gamma}{\l - \f12}\right)^{\f12}\, \int d t\,  d^2 x  \sqrt{\hat g}\, h\, \left\{ - \frac{1}{\sqrt{\hat g}}\, \p_t \sqrt{\hat g}\, \p_t   + (\hat D^2 + \hat R)^2 \right\}\, h \, ,
\ee
where $\hat D$ is the spatial covariant derivate constructed from the auxiliary metric $\hat g_{ij}$, and $\hat R$ the associated curvature.
The coefficient $(\g/(\l -\f12))^{1/2}$ in front of the integral decouples when taking the logarithm of the second functional derivative, giving an ultra-local contribution which can then be discarded. We thus are left with the operator
\be \label{hatD}
\mathcal{\hat D} = - \frac{1}{\sqrt{\hat g}}\, \p_t \sqrt{\hat g}\, \p_t   + (\hat D^2 + \hat R)^2 \, ,
\ee
for which we can use the results of  \cite{Baggio:2011ha}, in combination with \cite{Gusynin:1988zt}, which we recall in App.~\ref{App:HK}.

From \cite{Baggio:2011ha} we can directly borrow the extrinsic curvature terms in $a_2$, as the $\hat R$ term in \eqref{hatD} cannot contribute to those. 
For the terms depending only on the Ricci scalar, we observe that the time derivatives cannot contribute to those and hence we can ad hoc choose a time-independent metric and use the standard results from \cite{Gusynin:1988zt}. Putting things together, we find
\be
\label{a2}
a_{2} = - \frac{1}{64\, \pi}\,  \left( \hat K_{ij}\, \hat K^{ij} - \f12\, \hat K^2\right) \,.
\ee
The coefficient \eqref{a2} does not contain powers of the expected $\hat R^2$ term, a result true for any operator of the  type $(D^2 + X)^2$ in $d=2$  \cite{Gusynin:1988zt},
and in agreement with the $X=0$ case of \cite{Baggio:2011ha}.
As a consequence, we can deduce that no renormalization of the overall coupling of $R^2$ will take place. 
Similarly using  \cite{Gusynin:1988zt}, as explained in App.~\ref{App:HK}, we also obtain
\be \label{a0a1}
a_0 = \f{1}{16\, \pi}\, , \;\;\; a_1 = \f{7}{48\, \pi^{3/2} } \hat R \, .
\ee

Plugging \eqref{a2} and \eqref{a0a1} into \eqref{trace-expansion} and integrating over the proper time we find
\be
\label{tracehat2}
\begin{split} 
\f12 \, \hat \Tr \ln(\mathcal{\hat D}) =& - \f12\, \int dt\, d^2x \sqrt{\hat g}\, \Big\{(\Lambda^4 - \m^4)\, \f{1}{16\,\pi} + (\Lambda^2 - \m^2) \, \f{14}{48\, \pi^{3/2} } \hat R \\
&-  \ln\left(\f\Lambda\mu\right)\, \frac{1}{16\, \pi} \left ( \hat K_{ij}\, \hat K^{ij} - \f12\, \hat K^2\right)  +  \mathcal{O}\left(\f{1}{\L^2}\right) \Big\} \,.
\end{split}
\ee
The only term of our interest is the logarithmic divergence, which we can now rewrite as
\be\label{s1loop}
S^{1-loop}_{log} = \frac{1}{32\, \pi}\, \left( \frac{\l - \f12}{\g} \right)^{\f12}\, \ln \left( \f\Lambda\mu \right)\,  \int dt\, d^2 x \sqrt{g}\, \left \{ K_{ij}\, K^{ij} - \f12\, K^2\right \} \, ,
\ee
having used \eqref{aux} to express it in terms of the original metric $g_{ij}$.

\subsection{Beta functions}
We can now reabsorb the logarithmic divergencies by rewriting the bare couplings as $g_{b,i} = g_{R, i} + \d g_i$, being $g_{b,i}$ the bare coupling of the local operator $O_i(x)$, $\d g_i$ a counterterm chosen so to cancel the divergences and $g_{R, i}$ the renormalized coupling.
More specifically, we define the renormalized couplings as
\be
\label{renormalized}
\begin{split}
\frac{2}{\k_R^2} & =   \frac{2}{\k^2} - \frac{1}{32\, \pi} \,  \left( \frac{\l - \f12}{\g}  \right)^{\f12} \,\ln \left( \f\Lambda\mu \right)\, ,\\
\frac{2\, \lambda_R}{\k_R^2} & =  \frac{2\, \lambda}{\k^2}   - \frac{1}{64\, \pi}\,  \left( \frac{\l - \f12}{\g}  \right)^{\f12} \,\ln \left( \f\Lambda\mu \right)\, , \\ 
\frac{2\, \g_R}{\k_R^2} & =  \frac{2\, \g}{\k^2} \, .
\end{split}
\ee
We can now solve the first of \eqref{renormalized} obtaining the expression of the renormalized coupling $\k^2$, which reads
\be\label{kren}
\k_R^2 = \frac{\k^2}{ \left( 1 - \frac{\k^2}{64\, \pi}\, \left( \frac{\l - \f12}{\g} \right)^{\f12}  \ln \left( \f\Lambda\mu \right) \,  \right)} =  \k^2 \, \left( 1 + \frac{\k^2}{64\, \pi}\, \left( \frac{\l - \f12}{\g} \right)^{\f12}  \ln \left( \f\Lambda\mu \right) \,  \right) + \mathcal{O}(\hslash^2) \, ,
\ee
which used back in \eqref{renormalized} leads to
\be\label{lgren}
\begin{split}
\l_R &= \l + \frac{1}{64\, \pi}\,  \frac{\k^2}{\g^{1/2}} \left( \l - \f12 \right)^{\f32}  \ln \left( \f\Lambda\mu \right) + \mathcal{O}(\hslash^2) \, , \\
\g_R &= \g\, \left(1 + \frac{\k^2}{64\, \pi}\, \left( \frac{\l - \f12}{\g} \right)^{\f12}  \ln \left( \f\Lambda\mu \right) \right) + \mathcal{O}(\hslash^2) \, .
\end{split}
\ee

The beta functions can be evaluated by stating the independence of the bare coupling from the renormalization scale $\mu$, i.e. $\mu\, \p_\mu\, g_b =\mu\, \p_\mu\, g_R + \mu\, \p_\mu\, \d g = 0$, 
which leads to the system of beta functions
\be\label{beta}
\begin{split}
\beta_{\k^2} = \mu\, \p_\mu \, \k_R^2 &= - \frac{\k^4}{64\, \pi} \left( \frac{\l - \f12}{\g} \right)^{\f12} \, ,\\
\beta_\l = \mu\, \p_\mu \, \l_R &=  \frac{\left( \l - \f12 \right)}{\k^2}\, \beta_{\k^2} \, , \\
\beta_\g = \mu\, \p_\mu \, \g_R &= \frac{\g}{\k^2} \, \beta_{\k^2} \, .
\end{split}
\ee
Since the right-hand side of \eqref{beta} are $\mathcal{O}(\hslash)$ we can substitute the bare couplings with the renormalized one everywhere in the beta functions.
Now we can use \eqref{beta} to find
\be
\mu\, \p_\mu \, \left(\frac{\l_R - \f12}{\g_R}\right) = \f{1}{\g_R}\, \b_{\l} - \frac{\l_R - \f12 }{\g_R^2}\, \b_{\g} = 0 \, ,
\ee
so that 
\be\label{constC}
\left(\frac{\l_R - \f12}{\g_R}\right) = b \, ,
\ee
being $b$ a constant. Inserting \eqref{constC} in the first of \eqref{beta} we can solve the differential equation for $\k_R^2$, obtaining
\be\label{kR}
k_R^2(\m) = \frac{64\, \pi}{b^{1/2}\, (\ln \f{\mu}{\m_0} + C)}\, , 
\ee
where $C$ is an integration constant fixed by the boundary condition at some initial scale $\m=\m_0$. 
Using  \eqref{constC} and \eqref{kR} in \eqref{beta} we can integrate the remaining two beta functions obtaining the flow of the renormalized couplings $\l_R$ and $\g_R$, which respectively read
\be\label{lR}
\l_R(\m) = \f12 + \frac{C_1}{\ln \f{\mu}{\m_0} + C} \, ,
\ee
\be\label{gR}
\g_R(\m) = \frac{C_2}{\ln \f{\mu}{\m_0} + C}  \, ,
\ee
being $C_1$ and $C_2$ other two integration constant. Moreover, inserting \eqref{lR} and \eqref{gR} in \eqref{constC} we can see that $b = C_1/C_2$.
 
We observe that the running coupling  \eqref{kR} has the standard behavior of an asymptotically free coupling, running to zero for $\m\to\infty$.
However, we note that also $\l_R-\f12$ and $\g_R$ have the same behavior, a fact which leads to a problem for the perturbative treatment of HL gravity.
We have argued before that the effective perturbative coupling is  $\epsilon$, and substituting \eqref{gR} and \eqref{lR} in \eqref{eps}, we find the renormalized coupling to be
\be
\epsilon_R^2 = \frac{\k_R^2}{\l_R - \f12} = \f{64\,\pi\,C_2^{1/2}}{C_1^{3/2}} \, , 
\ee
so that it does not run to zero in the ultraviolet limit, but instead it remains constant along the renormalization group flow.
That is, the coupling $\epsilon$ is marginal at one-loop order.
Since the parameter $\epsilon$ characterizes the interaction strength of the theory, we are then in a situation in which the strength of the interaction remains finite at all scales, in particular meaning that the theory is not asymptotically free.

\section{Conclusions}
\label{Sec:concl}
We have presented here a one-loop calculation in the context of Ho\v rava-Lifshitz gravity. Due to the complexity of the full theory we restricted our analysis to a toy model, namely the conformal reduction of the projectable theory in 2+1 dimensions. For this particular choice of the dimension the conformal mode of the spatial metric is the only degree of freedom of the theory, thus we expect that the conformal reduction captures the main qualitative features of the model.
We have evaluated the renormalization group flow at a one-loop level for the dimensionless couplings of the model, in order to better understand the UV properties of such type of theories, and in particular to assess whether they can be asymptotically free.
Although we found that Newton's constant runs towards zero value in the UV, we also discovered that the coupling $\l$ flows to one half as fast as the Newton constant, implying that the perturbative parameter $\epsilon$ remains finite at all scales, thus spoiling the hopes of asymptotic freedom of the theory.

Looking back at \eqref{tot-var}, we can interpret the origin of such situation as a competition between the would-be asymptotic freedom of Newton's constant, and the strong coupling phenomenon that occurs when approaching $\l=1/2$. The latter is indeed a singular limit, in which the scalar mode is non-propagating. A similar strong-coupling phenomenon was pointed out in \cite{Charmousis:2009tc} in relation to the supposed IR limit $\l\to 1$ of the full HL theory, and it can be generically expected that some form of strong coupling or discontinuity will be associated to the disappearance of degrees of freedom due to enhanced symmetry, as for example in the massless limit of gravitons \cite{vanDam:1970vg,Zakharov:1970cc}.
In our case, the enhanced symmetry could be traced back to an anisotropic version of Weyl invariance at $\l=1/2$ and $\g=0$ \cite{Horava:2008ih}.
In analogy to the isotropic case, where scale invariance and unitarity of a quantum field theory imply conformal invariance (up to anomalies) in two \cite{Polchinski:1987dy} and seemingly four dimensions \cite{Dymarsky:2013pqa,Farnsworth:2013osa},
we might expect to have anisotropic Weyl invariance at a fixed point of the renormalization group in HL gravity (again up to anomalies \cite{Adam:2009gq,Baggio:2011ha,Griffin:2011xs}), and we can thus conjecture that our conclusion will apply also to the full theory, at least for what concerns $\l$. 
As we have restricted our theory to the projectable case, we miss the necessary terms to make the spatial part Weyl invariant \cite{Griffin:2011xs},
but  anisotropic Weyl invariance could be realized at a  fixed point with $\g\neq 0$ for the non-projectable model.

We should emphasize that while in \cite{Horava:2008ih,Horava:2009uw} a two-parameter family of fixed points was correctly identified, what we found here means that only one of them is reached by the interacting theory.
In order to better explain such point, it might be useful to look at a similar situation, by recalling what happens for a massless scalar field theory in four-dimensional curved spacetime with non-minimal coupling  $\x\, R\, \phi^2$.  
Being quadratic in the scalar field, we could include the non-minimal coupling term in the free action, and as $\xi$ is dimensionless we deduce that it defines a one-parameter family of fixed points.
However, the beta function for the quartic self-interaction coupling $g$ and the coupling $\x$ in the $\overline{MS}$-scheme read respectively \cite{Buchbinder:1989zz,deBerredoPeixoto:2003hv}
\be\label{scalarAF}
\b_{g} = \frac{3\, g^2}{(4 \,\pi)^2} \, , \hspace{1cm} \b_\x = \frac{g}{(4\, \pi)^2}\left( \xi - \f16\right) \, ,
\ee
and integrating them from a negative initial condition for the coupling $g$ (so that it runs to zero in the UV limit, instead of hitting a Landau pole) we find that $\xi(\m)\to 1/6$ for $\m\to\infty$, independently on the initial value $g(\m_0)<0$.
In this case $\xi=1/6$ is the value at which the theory shows conformal invariance at the classical level, and so analogously to our situation it is a value which is preferred by the flow trajectories, being the only one among the line of Gaussian fixed points that can be reached by the interacting theory. Of course the analogy is limited to this observation, the scalar theory being truly asymptotically free (albeit unbounded from below), and not loosing any degree of freedom as a consequence of Weyl invariance.

For completeness, we should point out that whereas for the reasons just discussed we expect the one-loop approach to the anisotropic Weyl invariant action to be a feature that the full theory will share with our toy model, we have no argument to support an analogous situation with the approach being such that the effective perturbative coupling $\eps$ remains finite. Furthermore, even in our toy model, $\eps$ might cease to be marginal at two loops or beyond. 
Only an explicit calculation could tell whether the additional degrees of freedom of the full higher dimensional model, or higher loop effects, might change the picture, however our toy model shows that potential troubles associated to strong coupling could be expected.

\newpage
\appendix
\section{The heat kernel coefficients}
\label{App:HK}

In order to keep this paper as self-contained as possible, we recall here the results of  \cite{Baggio:2011ha} and \cite{Gusynin:1988zt}.

In \cite{Baggio:2011ha}, the authors studied the scalar theory described by the following action
\be\label{BaggioAction}
S[\phi; g_{ij}]  =  
\frac{1}{2} \int d t\,  d^2 x  \sqrt{\hat g} \, N \phi\, \mathcal{\hat D}_0 \, \phi   \, ,
\ee
where
\be\label{BaggioOp}
\mathcal{\hat D}_0 =  - \frac{1}{N\, \sqrt{\hat g}}\, \p_t \frac{1}{N}\,\sqrt{\hat g}\, \p_t   + \f{1}{N} \hat D^2 N\hat D^2 \, ,
\ee
and in order to find the associated conformal anomaly, they computed the first three coefficients in the heat kernel expansion \eqref{HK-exp} for the operator $\mathcal{\hat D}_0$, thus finding
\be\label{Baggio-a0a1}
a_0 = \f{1}{16\, \pi}\, , \;\;\; a_1 = \f{1}{48\, \pi^{3/2} } \hat R \, ,
\ee
\be\label{Baggio-a2}
a_{2} = - \frac{1}{64\, \pi}\,  \left( \hat K_{ij}\, \hat K^{ij} - \f12\, \hat K^2\right) \, .
\ee

As explained in the text, our operator \eqref{hatD} differs from $\eqref{BaggioOp}$ in the spatial part, but in order to compute the effect of that, we can choose a time-independent background and exploit the results of \cite{Gusynin:1988zt}, where the first three non-zero heat kernel coefficients for the scalar operator
\be
\cD_2 = (- g^{\mu\nu}\, \na_\mu\, \na_\nu)^2 + V^{\mu\nu}\, \na_\mu\, \na_\nu + B^\mu\, \na_\mu + X\, ,
\ee
were computed on a general $d$-dimensional manifold, and arbitrary tensors $V^{\mu\nu}$, $B^\m$ and $X$.
In this case the heat kernel expansion writes
\be \label{HK-Gusynin}
\mathcal{H}(x,x,s; \mathcal{D}_2) = \sum_{n=0}^{\infty} \, s^{\frac{n-d}{4}}\, E_n(x;\mathcal{D}_2) \, ,
\ee
The result reads
\bea\label{higherorders}
&\hspace{-0.5cm}E_0(x; \cD_2) = \frac{1}{(4 \pi)^{\frac{d}{2}}}\,&\hspace{-0.5cm} \frac{\Gamma(\frac{d}{4})}{2\, \Gamma(\frac{d}{2})}\, ,\nonumber\\[2mm]
&\hspace{-0.5cm}E_2(x; \cD_2) = \frac{1}{(4 \pi)^{\frac{d}{2}}}\,&\hspace{-0.5cm} \frac{\Gamma\bigl(\tfrac{d-2}{4}\bigr)}{2\, \Gamma\bigl( \tfrac{d-2}{2} \bigr)}\, \left\{\f16\, R+ \f{1}{2d}\, V\right\}\, ,\\[2mm]
&\hspace{-0.5cm}E_4(x; \cD_2) = \frac{1}{(4 \pi)^{\frac{d}{2}}}\,&\hspace{-0.5cm} \frac{\Gamma(1 + \f{d}{4})}{2\, \Gamma(1 + \f{d}{2})}\, \Big\{   (d-2)\, \left (\frac{1}{90}\, R^{\a\b\g\d}R_{\a\b\g\d} - \frac{1}{90}\, R^{\a\b}R_{\a\b} +
\frac{1}{36}\, R^2 + \frac{1}{15}\, \na^2 R \right) \nonumber \\
&&\hspace{1.5cm} + \frac{d+4}{6\,(d+2)}\, \na^2V - \frac{2\,(d+1)}{3\,(d+2)}\, \na^\a \na^\b V_{(\a\b)} + \frac{1}{4\,(d+2)}\, V^2 + \nonumber\\
&&\hspace{1.5cm} + \frac{1}{2\,(d+2)}\, V^{(\a\b)}V_{(\a\b)} + \frac{1}{6}\, V\,R - \f13\,V^{(\a\b)}R_{\a\b} + \na^\a \, B_\a - 2\, X  \Big\}\, . \nonumber
\eea
where $V = V_\a{}^{\a}$ and 
\be
V^{(\a\b)} = \f12 (V^{\a\b} + V^{\b\a}) \, . 
\ee

For our purposes we need to specialize \eqref{higherorders} to $d=2$, interpret $\na_\a$ as $\hat D_\a$, take
\be
V^{\m\n} = 2 \hat g^{\m\n} \hat R\, , \;\;\; B^\m=0\, , \;\;\; X=\hat R^2\, ,
\ee
and multiply by an extra factor $(4\pi)^{-1/2}$ because of the extra (time) dimension in the trace.
As a result we get
\be\label{tindep-a0a1a2}
a_0 = \f{1}{16\, \pi}\, , \;\;\; a_1 = \f{7}{48\, \pi^{3/2} } \hat R \, , \;\;\; a_{2} = 0 \, ,
\ee
which is the time-independent (vanishing extrinsic curvature) version of the heat kernel coefficients we need for our one loop computation.
Combining \eqref{tindep-a0a1a2} with \eqref{Baggio-a2} we obtain the full coefficients \eqref{a0a1} and \eqref{a2}.

\providecommand{\href}[2]{#2}\begingroup\raggedright\endgroup


\end{document}